\documentstyle [12pt, russian] {article}
\begin{document}
\large
\centerline{\bf Duality in process of noncommutative deformation}
\centerline{\bf and topological nature of integral Cherepanov-Rice}
\centerline{\it Trinh V.K}
\centerline{\it Department of physics of polymers and crystals, MGU}
\centerline{\it E-mail : khoa@polly.phys.msu.ru}

Summary:  In this papers it is showed, that for the noncommutative
          deformation 
          simultaneously exist also loading deformation $H $,
          and unloading deformation
          $H^v $. The real deformation is a combination of these types
          deformations.
          The criterion of destruction $J $ reflects topological
         character of environment,
          i.e. defines properties of symmetry of
           environment at destruction. It is possible
          to tell, that during destruction the energy is released not
         continuously, and discretely. This situation is reflected
          through
           topological number Q or number  of unloading, connected to him.

1. As is known, after removal of loading an elastic body always
comes back in an initial status. The given definition of
elasticity
is a little simplified. If the pressure(voltage) exceeds a limit
of elasticity,
the dependence between loading and deformation ceases to be
 one-digit and depends on the order of the application of
loading [1]. For  simplicity, it is
  supposed, that
the equation of a curve is

$$ \sigma = \Phi (e) $$
fair at any program of the application of load, when a stress
monotonously grows. Nevertheless for real materials the business
is
more complicated. Let we have finished loading up to a point
$ A = (\sigma ^ {'}
, e ^ {'} $ of areas of plasticity. After that is made
unloading. Then the stress $ \sigma $ decreases up to zero. In
this process it is reflected not only plastic behaviour of
material, but also elastic. More careful experiments have shown,
that the law of unloading is not described precisely by direct
 line. Replacing it
by its closest straight line, we find, that its
inclination not
corresponds(meets) in accuracy to the initial module of
elasticity. At
polymeric materials, and also at composite  materials,
for example, fibreglasses, the law of unloading differs from
the law
Hooke  very essentially. Apparently, at the application of load
of material in
it the defects are formed. Naturally there is a question:
Whether exists duality of statuses of loading and unloading? And
in general, whether exists duality during plastic
deformations? Let's consider this problem in given
 clause.

2. As is known, under action of loading material  passes in
plastic status. All data of experiments show, that in
this status a the stress and the deformation is strongly
in fluctuation. More
that, diagram stress - deformation $ \sigma=f (e) $ is similar
 with
 the diagram pressure - volume $P=f (V) $ in a liquid [2].
 On the other hand
, the local phase transition in the end of a crack is observed
[3-5]. Thus, status,
close to destruction it is possible to consider(count) as a
status of transition or
critical, in which spontaneously arises a structure [6].
Density of distribution of probability of transition to plasticity
is the solution of the equation Fokker - Panck. However, using
potential character of process of deformation, instead of the equation
Fokker - Planck we can use the equation Schrodinger. It
means, that it is possible to consider(count) process of plastic
deformation
by process with supersymmetry. Here tensor of deformation  is
considered
in the role of parameter of order. In work [1] A.A. Iliushin has
presented on
discussion the concept of a trajectory of deformation in
space $E_5 $.
Actually, the space $E_5 $ is a layer in stratified
space of deformation [6]. It means, that a field of deformation
 we
 can consider(count) as a gauge field, in which generalized
 trajectory of
deformations Iliushin is the section of a stratification of
deformation. At
identification of the generalized trajectory of deformation
with knots we
have shown topological character of process of deformation
connected with
invariant Witten. Thus, density of
distributions of probability of transition to plasticity is
received[6] :

$$
Z = \int\exp [-kS _ {cs} (e)] {\cal D} e,
$$

 Here $k- $ it is factor
elasticity, $S _ {cs} - $ action Chern - Simons

 $$
S _ {cs} = \int\Gamma _ {cs},
$$

 where $ \Gamma _ {cs} $ - the form Chern - Simons
 for variety of deformation.

3. The spontaneous infringement of symmetry has allowed us
to receive
laminar  structure in process of thermomechanical deformation [6].
Here from this mechanism we can describe two processes of loading
 and
the unloadings with their topological character, which will be
considered
by dual processes. Processes of deformation in the end of a line
dislocation or the cracks are those examples of unloading, which
we would like to consider. These processes of clearing of energy
of the unloadings were used in model of seismology. Nevertheless
it is necessary to notice, that the nature of process of plastic
deformation -
it is noncommutative nature. Thus, the process of deformation is
 given
$P = \{ E_{ij} (x, t) \} $. It is noncommutative
space of deformation.  The deformation wave is the noncommutative
 wave. Besides it is possible to consider a scalar field
 describing
actions of defect on deformation [7]. For the complete description
of the distributions  of a deformation wave it is necessary to us
to use the
space being tensor product $C(R_4) \otimes M_n $, and 
$C(R_4) $ - algebra of smooth functions determined on
usual space - time, and $M_n $- algebra $n \times n $ of matrixes.
Now our task is a construction of noncommutative bundles
 of deformations above $C(R_4) \otimes M_n $. This model
of deformation was
 constructed in [6]. So, we admit(allow) connection as $ \omega =
A_{\mu} dx^{\mu} + (B_k-iE_k\otimes 1) \theta^k $, where
$A_{\mu} $
is function in $V $, receiving value in $M_n (C) \times
M_r (C) $, $ \mu\in \{0,1.., s \} $, ~ $B_k $- function above
$V $, having
the value in $M_n (C) \times M_r (C) $. Thus, action will be
 expressed in the form:

  $$
\begin {array} {l}
S = -\int\limits _ {R ^ {s+1}} \frac {1} {4n} Tr (F _ {\mu\nu} F ^ {\mu\nu}) + \frac {1} {2n}
Tr [(\nabla _ {\lambda} \phi_k) (\nabla ^ {\lambda} \phi^k)] \\
\qquad {} + \frac {1} {4n}
\sum Tr [[\phi_k, \phi_l] - {\bf m} \sum C _ {klm} \phi_m] ^2
\end {array}
$$

Where
$F ^ {a} _ {\mu\nu} = \partial _ {\mu} A _ {\nu} -\partial _ {\nu} A _ {\mu} +
[A _ {\mu}, A _ {\nu}] $,
$ \phi_k = {\bf m} B_k $,
$ \nabla _ {\lambda} \phi_k =\partial _ {\lambda} \phi _ {k} +
[A _ {\lambda}, \phi _ {k}] $

 " Magnetic monopol ", arising during evolution gauge
fields, actually is
 the deformation of the unloading. Really, from action
$ S $ we shall receive the equation of movement

$$\delta S=0 $$

This is nonlinear
 equation, his(its) solution with final energy is important
 for research of the
topological character of system. It is wonderful, that,
 not solving this
equation, it is possible to receive some information on
its properties.
 For this purpose from Lagrangian it  is simply necessary
 to find expression for
retentive energy of system, which for static decisions, with
 the account that $A ^ {a} _ {0} =0 $, looks like

$$ \begin {array} {l}
W = -\int\limits _ {R ^ {s}} d ^ {s} x\frac {1} {4n} Tr (F _ {\mu\nu}
 F ^ {\mu\nu}) + \frac {1} {2n}
Tr [(\nabla _ {\lambda} \phi_k) (\nabla ^ {\lambda} \phi^k)] \\
\qquad {} + \frac {1} {4n}
\sum Tr [[\phi_k, \phi_l] - {\bf m} \sum C _ {klm} \phi_m] ^2
\end {array} $$

This integral looks like integral Cherepanov - Rice without
kinetic energy. Energy achieves minimum and addresses in
zero at

 $$
A _ {0} (x) =0 ~, ~\phi _ {i} (x) = \frac {1} {2} {\bf m} E_i ~,
 ~\nabla _ {\lambda} \phi^i=0
$$

 Last equation turns to the equation $ \partial_i\phi^a=0 $.
However, in this case there are two gauge orbit. Let
$ \Omega_0 $- vacuum appropriate to an orbit
$ (A _ {\mu} =0 ~, ~\phi_k=0) $, and
$ \Omega_1 $- vacuum appropriate
to orbit $ (A _ {\mu} =0 ~, ~\phi_k=i {\bf m} E_k) $. Thus,
this condition is carried out for each orbit. In a case,
 when it is required for
finiteness of energy $W < \infty $, then there is a
following condition:
because of that the condition of finiteness $W $ consists
in enough
fast aspiration of fields to some configuration with $W=0 $ on
spatial infinity, it is visible, that the condition for
 $\phi$
at $r\equiv \mid x \mid \to \infty $ will look like
$r ^ {3/2} \nabla_i\phi \to 0 ~, ~\phi\phi \to F $. For getting
character of the field $A $ we shall present $ \nabla_i $ in
spherical coordinates
$ \{r, \theta, \phi \} $, $ \theta $- the component is
determined by expression
$ (\nabla
\phi^a) _ {\theta} = \frac {1} {r} \frac {\partial\phi^a}
 {\partial\theta} +
gC ^ {abc} A^b _ {\theta} \phi^c $. For performance of the
 condition given
above, in the field $A^b _ {\theta} $
 to be satisfied condition
$A ^ {b} _ {\theta} = \frac {1} {r} A ({\bf n}) $,
 where $A ({\bf n}) $
depends only on a spatial corner: $ \sum n_i^2=1 $.
 Other components have
 similar character. Thus, there are available
boundary conditions in the infinity

$$
A ^ {a} _ {\mu} = \frac {1} {r} A ({\bf n}) ~,
 ~ \phi^a =\frac {1} {2} {\bf
m} E_a ~, ~ r ^ {3/2} \nabla_i\phi = 0
$$

So, shall receive topological index, which was similar
expression in
commutative case

$$ Q =\int d^3 x K_0 $$

and topological
 current looks like

 $$ K _ {\mu} = \frac {1} {8\pi} Tr (\partial ^ {\mu} \hat
{\phi^a} \partial ^ {\varrho} \hat \phi^b)
\partial ^ {\sigma} \hat
\phi^c $$

Where $ \hat \phi_a =\phi^a / {\mid \phi\mid} ~,
 ~ \mid \phi\mid =
[ Tr (\phi\phi)] ^ {1/2} $. The charge of unloading looks like
$m =\int\frac {K_0} {g} d^3 x=Q/g $. In a case $n, r=1 $ we
 shall receive
usual monopol t Hoof- Polyakov.

4. Let deformation environment has symmetry $G $. For
simplicity
let's assume, that $G $- it is semisimple group. In process of
deformations the spontaneous infringement of symmetry
$G \to H\subset G $ will appear
. Our purpose - to investigate structure of a component
$A (\theta) $. In
commutative case in [8] approved, that a component
 $A (\theta) $ of
monopole is full determined through the Higgs field  $ \phi $.
In ours
noncommutative  case on the basis of the equation of movement
we also
 receive the same statement. From the requirement of finiteness
  of energy
in a stationary status or from integral such as Cherepanov - Rice
 we
 receive topological invariant. Shall copy this result in
 the equivalent form, which does not diminish a generality of
  the problem:

 $$\exp (4\pi i\epsilon
A (\theta)) =1 
$$
 where $ \epsilon $- a constant of connection of defect with
 environment.
Let $ {\cal L} (H) $ - Lie algebra of group $H $, and
 $T_1, T_2... T_r $
 its generators. Thus, under the theory of Lie group and Lie
 algebra
[10] it is possible to find gauge maps $S\in H $,
so that

$$\epsilon A (\theta) =S\sum\limits _ {i=1} ^ {r} \beta_i T_i
S ^ {-1} $$

 The system $ \beta_1, \beta_2 ...\beta_r $- is named as weight
of statuses of unloading. Using this concept, it is possible
to copy
condition of "quantization" as

$$\exp (4\pi
i\sum\limits _ {i=1} ^ {r} \beta_iT_i) =1 $$

If $H $ is group
for loading, it is necessary to search for group $H ^ {v} $
for unloading.
Thus, our purpose is to deform system of weight of group
 $H ^ {v} $,
satisfying to the condition above mentioned. For this purpose
 we use method in
[9]. Let for group $H $ there is a system of roots
$ \alpha \in \Phi (H) $. Then system of roots $ \omega $ of
universal enveloping  group $ \tilde H $ will look like

$$
\Lambda (\tilde H) = \{ \omega ~; ~ 2\omega.\frac{\alpha}{\alpha^2}
\in Z ~; ~ \alpha\in
\Phi (H)  
$$

and $ \Phi (H) $ - system of roots of group $H $. First stage
given in
[9] is to deform system $ \beta $ so that
the condition was satisfied :

$$
\widetilde {\exp} (4\pi i\beta. T) \in
Z (\tilde H) 
$$
where the centre - $Z (\tilde H) $. Here
there is a system of dual roots
$ \Phi ^ {v} (H) = \{\alpha^v=N ^ {-1} \frac{\alpha}{\alpha^2} ~;
 ~ \alpha\in \Phi (H) \} $. 
For the elementary case, when $H $- simple group,
$N $ will be number as
$N^2 = (\sum\frac{1}{\alpha^2}) /\sum\alpha^2 = $.
System of weight
$ \Lambda ({\tilde H} ^v) $ 
has the form 
$ \Lambda ({\tilde H}^v) = 
\{ \omega ~; ~ 2\omega N\alpha\in Z ~; ~\alpha\in \Phi (H) \} $.
 The greatest system of weight $ \beta $,
satisfying
condition $2\alpha.\beta\in Z ~, ~\alpha\in \Phi (H) $,
is written as
$ \beta=N\omega ~, ~\omega\in \Lambda ({\tilde H} ^v) $.
 The following
 step is to look for system carrying out a condition
 $ \exp (4\pi
i\sum\beta_i T_i) =1 $ or condition $2\beta.\omega\in
Z ~, ~\omega\in \Lambda (H) $. The general form of
system $ \beta $,
carrying out the described above condition, looks like
 $N ^ {-1} \beta \in
\Lambda (H^v) $. System $ \beta $
determines all characters of group $H^v $

5. So, process of destruction of environment the following.
A crack (defect) is given
in environment with group of symmetry $G $.
Under action of loading system of
environments with a crack passes in a status of destruction
(status of
phase transition). Thus there is a spontaneous infringement
symmetry $G\to H $. More precisely, it is the process
$G\to H\otimes H^v $. It
means, that simultaneously exist also loading $H $,
 and unloading
$H^v $. The real deformation is a combination of these types
deformations. The presence of topological objects creates a
difference
between loading and unloading. From the requirement of
finiteness of the energy of
deformation and under influence of spontaneous infringement
of symmetry for
stationary status we shall receive, that integral of energy
is invariant. Simultaneously the expression under integral is
 potential of deformation. Thus, this integral looks like
 integral
Cherepanov - Rice $J $ without kinetic parts  of energy. In
 the classical mechanics of destruction integral
 Cherepanov - Rice
 is considered as criterion of destruction. Actually,
 on our model,
the criterion of destruction $J $ reflects topological
character of environment,
i.e. defines properties of symmetry of
environment at destruction. It is possible
to tell, that during destruction the energy is released not
continuously, and discretely. This situation is reflected
through
topological number Q or number  of unloading, connected to him

\end {document}